# Voltage Gated Domain Wall Magnetic Tunnel Junction-based Spiking Convolutional Neural Network

Aijaz H Lone[1], Hanrui Li[1], Nazek El-Atab[1], Xiaohang Li[1,2] and Hossein Fariborzi[1]

[1]Computer, Electrical, and Mathematical Sciences and Engineering (CEMSE) Division, KAUST,
[2]Physical Science and Engineering Division, KAUST

*Abstract:* We propose a novel spin-orbit torque (SOT) driven and voltage-gated domain wall motion (DWM)-based MTJ device and its application in neuromorphic computing. We show that by utilizing the voltage-controlled gating effect on the DWM, the access transistor can be eliminated. The device provides more control over individual synapse writing and shows highly linear synaptic behavior. The linearity dependence on material parameters such as DMI and temperature is evaluated for real-environment performance analysis. Furthermore, using skyrmion-based leaky integrate and fire neuron model, we implement the spiking convolutional neural network for pattern recognition applications on the CIFAR-10 data set. The accuracy of the device is above 85%, proving its applicability in SNN.

## I. INTRODUCTION

Neuromorphic computing takes its inspiration from the energy-efficient computational power of the brain [1]. The realization of neural network accelerators using CMOS devices is limited by the high energy cost associated with the Von-Neumann Bottleneck [2][3]. To mitigate this problem many alternative solutions such as memristor-based hardware implementation of the neural networks have been investigated [4][5]. Spintronic devices can perform energy-efficient logic, processing, storage, and sensing in a single device [6][7][8][9]. Due to these merits, spintronic devices are prime candidates as synapses and neurons for neuromorphic computing implementation [10][11][12][13]. In particular, magnetic domain wall motion (DWM)-based devices have gained substantial interest in the neuromorphic computing community [14][15][16]. These devices, when driven by the spin-orbit torque, show synaptic capabilities with better linearity compared to other memristors [17][18][19][20]. While most of these devices are introduced and evaluated based on ideal models and simulations, for an actual circuit implementation an access transistor is required to have control over individual synapses and to avoid writing errors caused by signals from other devices in the crossbar [21][22]. We propose a novel DWM-based MTJ synapse, driven by SOT and gated by the voltage at two ends, to prevent DW annihilation to edge. The proposed scheme of temporary DW pinning at gates can act as a solution to minimize the writing error and mandatory access transistor requirement in a 1T1R crossbar configuration. The voltage control of the magnetic anisotropy (VCMA)-based spintronic devices has been presented as a more energy-efficient switching solution [23][24][25]. The proposed device shows highly linear synaptic behavior. We have evaluated the linearity dependence on material chiral Dzyaloshinskii Moriya interaction (DMI) and temperature for studying device performance in the real environment.

Furthermore, the biologically inspired, spiking neural networks (SNNs) can mimic the operation of the human brain and have shown superior energy efficiency and performance in many applications [26][11][27]. Thus, hardware for SNNhas become hot research topic over several years. The information in SNNs is conveyed by spike, in terms of the spike rates, latencies, and possibilities, which shows the good property in low resource utilization and energy-efficient information processing system. In conjunction with the skyrmion based leaky integrate and fire neuron model developed in our previous work, we implement the spiking convolutional neural network for pattern recognition applications based on the proposed DW-MTJ device. The DW-MTJ resistance is used as the weights in SCNN and when trained and tested on CIFAR-10 data set, device shows 85% accuracy in image recognition tasks.

## II. DEVICE MODELING AND SIMULATIONS

The micromagnetic simulations of the proposed device are carried out using the micromagnetic simulator MuMax3 [28]. The Landau–Lipschitz–Gilbert (LLG) equation with custom field SOT terms is used to study the magnetization time-evolution and domain wall motion:

$$\frac{d\widehat{m}}{dt} = \frac{-\gamma}{1+\alpha^2}[\widehat{m} \times \overrightarrow{H_{eff}} + \widehat{m} \times (\widehat{m} \times \overrightarrow{H_{eff}})] + \overrightarrow{\tau_{SOT}} \quad (1)$$

where $\widehat{m}$ is the normalized magnetization vector, $\gamma$ is the gyromagnetic ratio, $\alpha$ is the Gilbert damping coefficient and $\overrightarrow{H_{eff}}$ is the effective field due to various magnetic energy terms.

$$\overrightarrow{H_{eff}} = \frac{-1}{\mu_0 M_S}\frac{\delta E}{\delta m} \quad (2)$$

The total magnetic energy of the free layer includes the exchange energy, Zeeman energy, uniaxial anisotropy energy, demagnetization energy, and DMI energy.

$$E(m) = \int_V [\, A(\nabla\widehat{m})^2 - \mu_0\widehat{m}.\overrightarrow{H_{eff}} - \frac{\mu_0}{2}\widehat{m}.\overrightarrow{H_d} - K_u(\widehat{u}.\widehat{m}) + \varepsilon_{DM}\,]dv \quad (3)$$

where A is the exchange stiffness, $\mu_0$ is the permeability, $K_u$ is the anisotropy energy density, $\overrightarrow{H_d}$ is the demagnetization field, and $\overrightarrow{H_{eff}}$ is the external field and $\varepsilon_{DM}$ is the DMI energy.

$\overrightarrow{\tau_{SOT}}$ is the SOT split into damping-like and field-like torque components [29].

$$\tau_{SOT} = -\frac{\gamma}{1+\alpha^2} a_J [(1+\xi\alpha)\hat{m} \times (\hat{m} \times \hat{p}) + (\xi - \alpha)\hat{m} \times \hat{p}]$$
(4)

The $a_J$ represents current density J and Hall angle $\theta_H$ as follows.

$$a_J = \left\| \frac{\hbar}{2M_s e} \frac{\theta_H}{d} J \right\| \quad (5a)$$

and

$$\hat{p} = sign(\theta_{SH}) \boldsymbol{j} \times \hat{n} \quad (5b)$$

The magnetization computed from the MuMAX serves as an input parameter for the non-equilibrium Green's function (NEGF)-based transport module which computes the device resistance [30][31]. Furthermore, the simplified synapse resistance model from our previous work [32] is modified as follows:

$$R_{SYN} = R_{AP} \frac{[1+\hat{m}\cdot\hat{m}_P]}{2} + R_P \frac{[1-\hat{m}\cdot\hat{m}_P]}{2} \quad (6)$$

where $R_{AP}$ and $R_P$ are the resistance in complete anti-parallel MTJ configuration and parallel configuration, respectively. $\hat{m}_P$ is the direction of reference layer magnetization.

## III. DOMAIN WALL MTJ DEVICE RESULTS AND DISCUSSION

In Fig. 1 we show the SOT-driven voltage-gated DWM- based MTJ device structure with different material layers. The writing into the device is performed by passing the charge current from WL to $\overline{WL}$ for depression and $\overline{WL}$ to WL for potentiation. The SOT generated by this charge current at the heavy metal (Ta) and free layer (CoFeB) interface moves the DWM smoothly across the magnetic free layer. The movement of the DW results in a change in the net free layer magnetization, which is detected by the active MTJ (Reference layer CoFeB/MgO/Free layer (area under MgO)). The device operation for different combinations of the current direction in Ta and gate voltages is as follows. Considering the device initially in depression, the potentiation in the device is realized by keeping Gate-2 (Vg2) at a positive voltage which lowers the anisotropy below Gate-2 and simultaneously current pulses are applied from $\overline{WL}$ to WL. The positive voltage Vg2 opens Gate-2 and also increases the DW velocity. Gate-1 is kept at a negative bias, which increases the anisotropy below the Gate-1 and as the DW reaches Gate-1, the DW speed reduces abruptly and the normalized magnetization value stabilizes around 0.6 as shown in Fig. 2(a). Fig. 2(b) shows the DW chirality switching from left-handed chirality (LHC) to right-handed chirality (RHC) as the DW approaches the high anisotropy barrier (Gate-1). Likewise, for the realization of the synaptic depression, Gate-1 is kept at a positive voltage while Gate-2 switches to a negative voltage. The current pulses are applied from WL to $\overline{WL}$ and the DW moves towards Gate 2 where it encounters a higher anisotropy barrier. The barrier stops the DW motion and prevents it from possible annihilation at the edge as shown in Fig. 1(b) considering normal DW- synapse the DW is lost to the edge but in presence of the gating scheme we show the DW stops before reaching the edge. Most importantly when the DW is between Pin1 and Vg1 the temporary increased anisotropy at site Vg1 doesn't allow DWM in the device. This scheme makes the proposed device less prone to writing errors caused by the current coming from other devices in the crossbar. Thus, the device can be used in a crossbar as a standalone unit without the need for a transistor which prevents a sneak path. In the synaptic depression phase, we observe the analog normalized magnetization switching from +0.6 to -0.6. Please Note that when considering only the region below the active region this magnetization change normalizes from -1 to +1 and the non-linearity of the device is about 0.04. At Gate 2 a DW chirality from RHC to LHC is observed as shown in Fig. 2(b). The DW chirality switching helps the SOT-driven DWM in the next writing cycle. Fig. 2(c) shows the free layer magnetization as a function of the DW-position for two consecutive cycles of synaptic operation. The magnetization increases linearly as a function of the DW-position till 200 nm after that a slight linearity degradation is observed due to the gating effect after 200 nm. This can completely be avoided by keeping gates at an optimum distance from the active MTJ. Furthermore, the operation in different phases is tabulated in Table II for more details. In all of these material stacks for SOT-driven DWM devices such as Ta/CoFeB, Pt/Co, and or Pt/CoFeB, the existence of DMI is ubiquitous. Here, we furthermore study the dependence of the DWM on this parameter. We observe that without DMI the DW doesn't move for even current density $J=8\times10^{11} A/m^2$. But in presence of a small DMI of value $D= 0.5 mJ/m^2$, the DW is tilted by a small angle and we observe the DW starts moving smoothly (steady motion) with velocity approaching 30m/s in 5 ns as shown in Fig. 3(a). The velocity increases as the DMI increases till $D= 1.4 mJ/m^2$ but on increasing the DMI above the threshold, DW velocity starts reducing. This is a clear indication of the Walker breakdown at higher DMI hence precessional DW- propagation. The average reduces from 40 m/s for $D= 1.4 mJ/m^2$ to around 20 m/s velocities for $D=2.2 mJ/m^2$ as shown in Fig. 3(a). Thus, for a reliable device operation, the optimum DMI value should range from (0.5 to 1.2) $mJ/m^2$. In Fig. 3(b) We show the MTJ resistance at different DMI values for 3 operating cycles. We observe irrespective of the Walker breakdown at 2 $mJ/m^2$ and 2.2 $mJ/m^2$ the device shows a highly linear resistance switching (plasticity) for both potentiation and depression. Compared to 2 $mJ/m^2$ and 2.2 $mJ/m^2$ the resistance switches fast for 1 $mJ/m^2$ as expected due to reduced DW- Velocity at higher DMI coefficients. The device linearity plays a crucial role in circuit and system-level neural network implementation. For an efficient weight realization, the synaptic device is expected to be ideally linear.

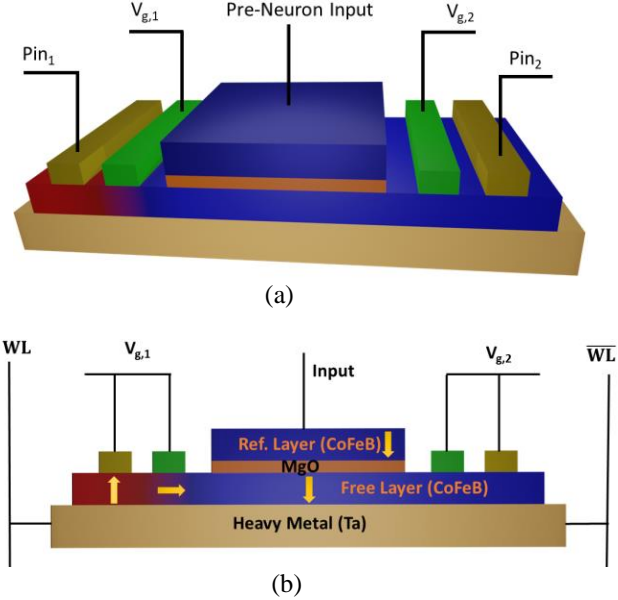

Fig. 1(a) SOT-driven voltage-gated DWM- based MTJ device structure. (b) Side view of the device showing writing and gating terminals.

Table I: Magnetic simulation details

| Grid Size | Cell Size (nm) | Anisotropy Ku J/m³ | Saturation Mag. Ms A/m | Exchange Stiffness J/m | DMI J/m² |
|---|---|---|---|---|---|
| 512,256, 0.9 | 1,1, 1 | 1×10⁶ | 1×10⁶ | 1×10⁻¹¹ | 1.0×10⁻¹² |

Table II: Device operation

| BL | $\overline{BL}$ | Gate-1 ($V_{g1}$) | Gate-2 ($V_{g2}$) | MTJ STATE | DW- Motion |
|---|---|---|---|---|---|
| 0 | 0 | 0 | 0 | Previous State | No |
| 0 | 1 | -V | +V | Potentiation | +X to -X |
| 1 | 0 | +V | -V | Depression | -X to +X |
| 1 | 1 | X | X | Previous State | No |

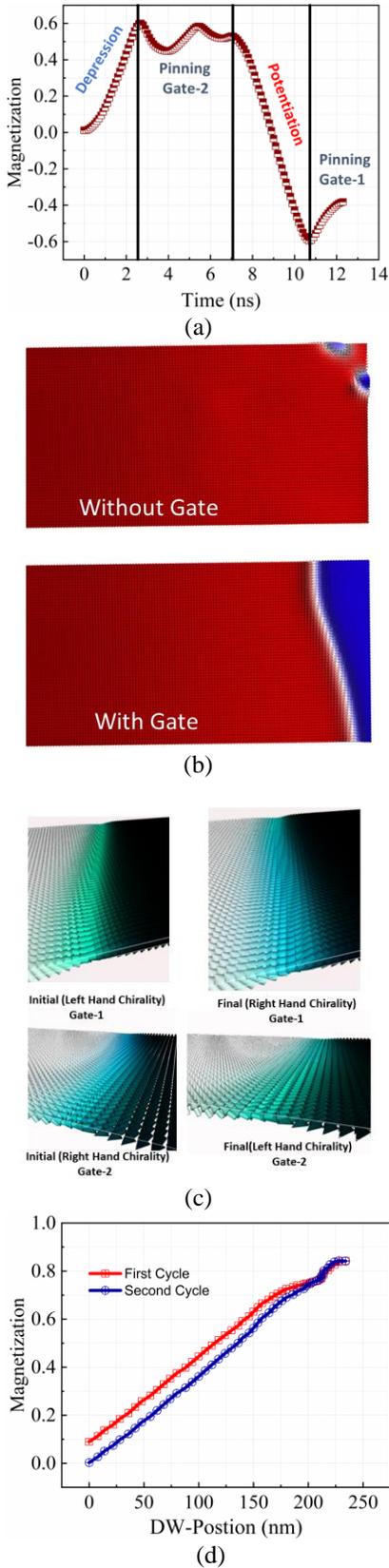

Fig. 2(a) DW- temporary pinning by gate bias and magnetization potentiation depression (b) Free layer magnetic texture showing annihilation of DW without gate and pinning when the gate bias is present (c) DW- chirality reversal at the pinning site and (d) Magnetization as a linear function of the DW position.

We compute the non-linearity (NL) of the device by adopting the following method for other charge-based memristors such as resistive RAM. The measured plasticity behavior is fitted to the non-linear weight behavioral model as shown in Fig. 3(c-d). The conductance increase/decrease is expressed as an exponential function of the number of pulses P and given by[33]:

$$G_{LTP}(V_g, V_d) = \beta\left[1 - e^{-(P/\alpha)}\right] + G_{min}(V_g, V_d) \quad (7)$$
$$G_{LTD}(V_g, V_d) = \beta\left[1 - e^{-\left(\frac{P-P_M}{\alpha}\right)}\right] + G_{max}(V_g, V_d) \quad (8)$$
$$\beta = \frac{(G_{max}(V_g,V_d) - G_{min}(V_g,V_d))}{\left(1 - e^{-\frac{P_M}{\alpha}}\right)} \quad (9)$$

Where, $G_{LTP}$ is the synaptic conductance, $G_{min}$ is the minimum conductance and $G_{max}$ is the maximum conductance achieved by the device, $\alpha$ is the non-linearity fitting parameter, $P_M$ is the maximum pulse number, $\beta$ is the function of $G_{min}$, $G_{max}$, $\alpha$ and $P_M$. Fig. 3(c) shows the simulated MTJ resistance in the potentiation and depression phases. The simulated resistance when fitted with the model gives an NL of around 0.04 which is almost the ideal case.

Furthermore, we use the above model to compute the NL of the DW synapse in presence of thermal noise. Fig. 3(d) shows the increasing NL of the DW- synaptic device due to thermal noise. As temperature increases the linearity decreases randomly such as NL = 0.31 for T = 200K, NL = 0.4 for T = 300K but for 360K the NL is 0.32. Although there is linearity degradation the device NL is still acceptable for recognition tasks.

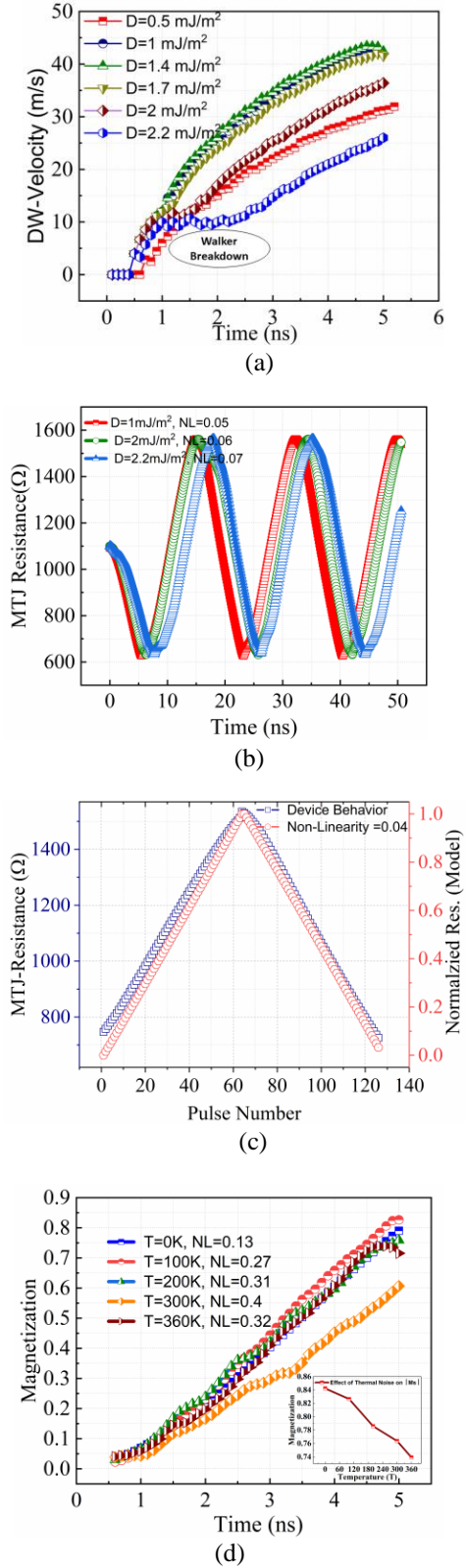

Fig. 3(a) DW-velocity for increasing DMI coefficients showing Walker breakdown after D = 1.4 mJ/m² (b) SW-MTJ resistance during potentiation and depression for different DMI coefficients (c)Model-based linearity fitting of simulated resistance and (d) Magnetization evolution determining the device behavior in presence of thermal effects.

## IV. SPIKING CONVOLUTIONAL NEURAL NETWORK (SCNN)

To verify the learning capabilities of MTJ-based synapses and leaky integrate and fire (LIF) neurons, we build a spike convolutional neural network (SCNN), which contains six spike convolutional layers (Conv) and two spike fully connected layers (SFC), as shown in Fig .4(b). We evaluate the model on the CIFAR-10 dataset [34], which is an open-source dataset containing 10 class images of 32x32 size. During training, the original image received from the dataset is first encoded towards time steps to generate discrete spikes input. As shown in Fig .4(a), the conductance values extracted from the MTJ device are used as the synaptic weights connecting neurons in different layers . For LIF neurons, the membrane potential $V(t)$ is updated for each time step and the leakages are set exponentially over time with the time constant as illustrated in Fig .4(a). When the membrane potential $V(t)$ is over the threshold, the neuron will generate a new spike and set it as $V_{rest}$. The skyrmion devices have demonstrated a similar property to work as LIF neurons [ref]. Together with the DW-MTJ-based synapse, after each convolutional layer or fully connected layer, we adopt the LIF neuron model directly from our device to mimic the biological neuron.

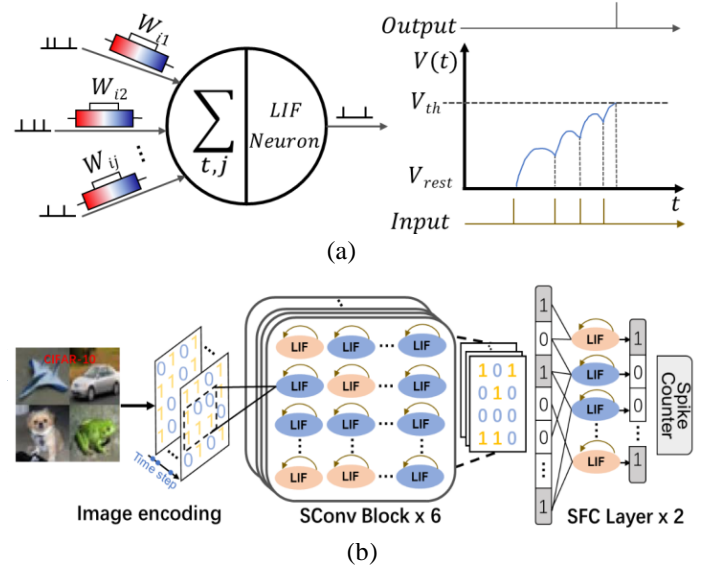

Fig. 4(a) The structure of the LIF neuron model, where our DW-MTJ-based device works as the synapse. (b) The illustration of the proposed SCNN model structure. The general model consists of three main parts, including an encoding layer, six spike-based convolutional blocks, and two spike-based fully connected layers.

The simulation results show that SCNN with MTJ-based synaptic device and skyrmion-MTJ-based quadratic LIF neuron shown in our previous work [35].
The skyrmion quadratic LIF neuron is modeled by

$$\frac{dx}{dt} = \frac{1}{k_1}(x^2 + k_2 x + \vartheta J) \quad (10)$$

$$\vartheta = -\left(\frac{\beta}{\alpha} + \frac{(\alpha-\beta)}{\alpha^3\left(\left(\frac{D}{G}\right)^2 + \alpha\right)}\right)\frac{Pa^3}{2eMs} \quad (11)$$

where $x$ is the position of the skyrmion for MTJ, $k_1$=2.12x10³ ms, $k_2$=-27.6 m, and $\vartheta = f(\alpha, \beta, G, D, P \text{ and } a)$ m³/C, with $\alpha$, $\beta$ is the non-adiabatic term, *G* as the gyromagnetic coupling, *D* is the dissipative force tensor, *P* is the polarization, *and* as the lattice constant, and *Ms* as the saturation magnetization. achieves comparable results with the software-based SCNN algorithm. The algorithm simulation results are illustrated in table III. When adopting the MTJ-based LIF neuron model in the network, the recognition accuracy receives approximately 78.53%, which is slightly lower than the ideal LIF neuron (85.43%). With both MTJ-based synapse and MTJ-based LIF neurons, the results indicate that the device with better linearity receives better recognition accuracy as expected. The competitive accuracy in image recognition and high linearity of the device indicate that the proposed device is highly suitable for SNN application for the inherent property of mimicking spike neurons and synapses.

Table. III: SNN Simulation results for combination of the proposed DW-MTJ synapse and different LIF neurons

| Method | Structure | Neuron | Synapse | Nonlinearity | Acc |
|---|---|---|---|---|---|
| CNN | 6Conv+2FC | / | / | / | 89.02 |
| SCNN | 6SConv+2SFC | Ideal LIF | Ideal | / | 87.56 |
|  |  |  | T=0K | 0.13 | 86.68 |
|  |  |  | T=100K | 0.27 | 85.13 |
|  |  |  | T=300K | 0.4 | 84.23 |
|  | 6SConv+2SFC | MTJ-based LIF | Ideal | / | 75.98 |
|  |  |  | T=0K | 0.13 | 70.04 |
|  |  |  | T=100K | 0.27 | 68.86 |
|  |  |  | T=300K | 0.4 | 61.16 |

## V. CONCLUSION

The SOT-driven and voltage-gated DW-MTJ device is presented for neuromorphic computing applications. The introduction of voltage control eliminates the writing error due to cross-talk. The device enables more control over individual synapse writing. The DWM is restricted to the region below the active region and we observe ideal linear synaptic behavior. Furthermore, using the skyrmion-based LIF neuron model, we implement an SCNN based on the proposed synaptic devices. The highly linear weights help in achieving 87.56% for ideal case88% accuracy.

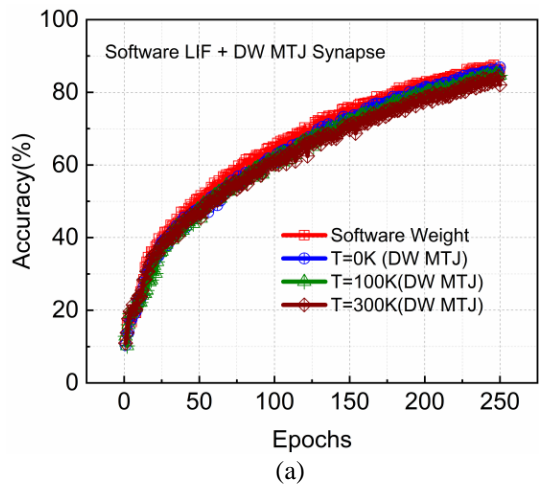

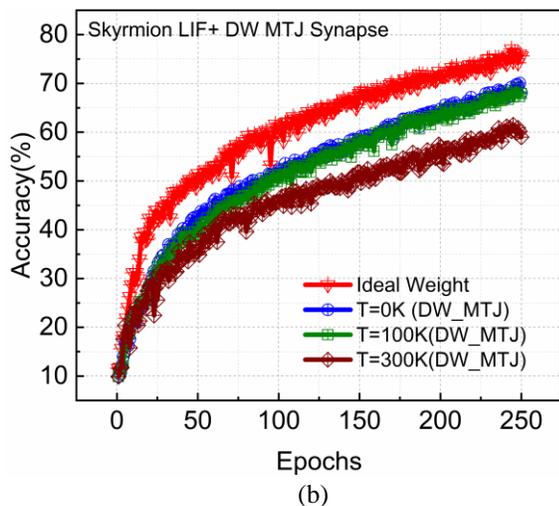

Fig. 5(a) DW-MTJ synapse device performance in terms of recognition accuracy in cooperation with the ideal neuron model and (b) Device performance with a skyrmion-based LIF neuron model.